\PassOptionsToPackage{square,comma,numbers,sort&compress}{natbib}
\documentclass[%
secnumarabic,graphics,nofootinbib,tightenlines,nobibnotes,floatfix,aps,
superscriptaddress,showkeys,preprintnumbers,amsmath,amssymb,preprint,
showpacs]{elsarticle}

\usepackage[english]{babel}
\usepackage{verbatim}
\usepackage{subfigure}
\usepackage{hyperref}
\usepackage{booktabs,caption}
\usepackage{float}
\usepackage{amsbsy} 
\usepackage{soul}
\usepackage[all]{xy}
\usepackage{tikz-cd}
\usepackage{color}
\usepackage{lineno}
\usepackage{amsmath}
\usepackage{amssymb}
\usepackage{slashed}
\usepackage{rotating}
\usepackage{dcolumn}
\usepackage{bm}
\usepackage{textcomp}

\def\be{\begin{equation}}
\def\ee{\end{equation}}
\def\bea{\begin{eqnarray}}
\def\eea{\end{eqnarray}}
\renewcommand{\d}{\mathrm{d}}

\begin{document}
\begin{frontmatter}
\title{Reparametrizations and metric structures in thermodynamic phase space}

\author[1]{V. Pineda-Reyes\corref{cor1}}
\ead{vpineda@ciencias.unam.mx} 

\author[1]{L. F. Escamilla-Herrera}
\ead{lenin.escamilla@correo.nucleares.unam.mx}

\author[2,3]{C. Gruber}
\ead{christine.gruber@uni-oldenburg.de} 

\author[4]{F. Nettel}
\ead{fnettel@ciencias.unam.mx}

\author[1,5,6]{H. Quevedo}
\ead{quevedo@nucleares.unam.mx}

\cortext[cor1]{Corresponding author}
\address[1]{Instituto de
	Ciencias Nucleares, Universidad Nacional Aut\'onoma de
	M\'exico,\\AP  70-543, Ciudad de M\'exico, 04510 Mexico}
	
\address[2]{Hanse-Wissenschaftskolleg Delmenhorst, Germany}
\address[3]{Institut f\"ur Physik, Universit\"at Oldenburg, 
	D-26111 Oldenburg, Germany}
	
\address[4]{Departamento de F\'\i sica, Fac. de Ciencias \\
	Universidad Nacional Aut\'onoma de
	M\'exico,\\AP  70-543, Ciudad de M\'exico, 04510 Mexico}
	
\address[5]{Dipartimento di Fisica and ICRA, Universit\`a di Roma 
	``La Sapienza", I-00185 Roma, Italy}
	
\address[6]{Institute of Experimental and Theoretical Physics, 
	Al-Farabi Kazakh National University, Almaty, Kazakhstan}

\date{\today}

\begin{abstract}
We investigate the consequences of reparametrizations in the geometric des\-crip\-tion of thermodynamics analyzing the effects on the thermodynamic phase space. It is known that the contact and Riemannian structures of the thermodynamic phase space are related to thermodynamic equilibrium and statistical fluctuations in the Boltzmann-Gibbs statistical mechanics. The physical motivation for this analysis rests upon the possibility of having, instead of a direct control of the intensive parameters determining the state of the corresponding physical reservoirs, the control of a set of differentiable functions of the original variables. Likewise, we consider a set of differentiable functions of the extensive variables accounting for the possibility of not having direct access to the original variables. We find that the effect of reparametrizations on the thermodynamic phase space can be codified, in geometrical terms, in its contact and Riemannian structures. In particular, we single out a rank-two tensor that enters in the definition of the metric which geometrically comprises the information about such reparametrizations. We notice that even if these geometric structures are modified by the reparametrizations, the metric structure on the thermodynamic space of equilibrium states is preserved. 
\end{abstract}
\begin{keyword}
Statistical Mechanics\sep Thermodynamics\sep Geometry
\end{keyword}

\end{frontmatter}

\section{Introduction}
\label{sec:int}

The use of a geometric language in thermodynamics started with the works of Gibbs \cite{1948Gibb} and Caratheodory \cite{1909Cara}, introducing the notion of the thermodynamic space of equilibrium states, or thermodynamic equilibrium space (TES), as a (hyper-)surface determined by a fundamental relation and the first law as a Pfaffian differential equation whose solutions form the TES. 
It was Hermann who first introduced the concept of a thermodynamic phase space (TPS) as a $(2n+1)$-dimensional manifold equipped with a contact structure, whose maximally integrable submanifolds constitute the TES \cite{1973Herm}. It is worth mentioning that, in the geometric description of thermodynamics, distinct manifolds can be considered, i.e., our definitions of the TPS and the TES do not necessarily coincide with those appearing in other works \cite{van_der_Schaft_2018}. In fact, from the point of view of thermodynamics we are in most of the cases interested in manifolds which are $\mathbb{R}^{2n+1}$. A different approach to thermodynamic geometry using Riemannian geometry was later proposed by Weinhold \cite{1975Weina} and Ruppeiner \cite{1979Rupp, 1995Rupp}. In these two approaches, a metric is defined on the TES. In the Weinhold geometry, the components of the metric are given by the Hessian of the internal energy, while in Ruppeiner's case by the negative Hessian of the entropy; later \cite{1984Salamon}, it was shown that these two metrics are related by a conformal factor and in \cite{conformalGauge} the same result is established in terms of a conformal gauge transformation in the TPS. In the two decades following the proposal of these two formalisms, there was a significant number of works devoted to the physical implications of these Riemannian structures in thermodynamics. In particular, several works were dedicated to the study of  the physical significance of the thermodynamic length, the relation to thermodynamic fluctuations and the link between the thermodynamic curvature, stability and phase transitions, among other related topics \cite{1979Rupp, 1983Salamon, 1984Salamon, 1985Salamon, 1990Janyszek, 1989Janyszek, Diosi1984}. 
Ref.\,\cite{1995Rupp} contains an extensive account of the different results concerning thermodynamic geometry, as well as a comprehensive collection of bibliographic references up to the date of its publication. More recently, there has also been established a parallelism between statistical inference theory, statistical mechanics and their geometric descriptions \cite{amari1982, Brody1995, amari2009methods,crooks}. 

In Ref.\cite{1990Mrug}, a statistical approach to thermodynamic geometry was explored by constructing a contact Riemannian manifold associated to a family of exponential probability distributions. Inspired by Jaynes' perspective on the interpretation of statistical mechanics as a theory of maximum entropy inference, that work establishes that the contact structure determines the thermodynamic equilibrium and that the metric structure is related to statistical fluctuations. Further explorations on the contact geometry of thermodynamics have followed \cite{MRUGALA1991109, MRUGALA1996339, 2006Salamon, MRUGALA1985197, RAJEEV2008768, paraSasakian, HamThermo, bravetti2017contact, conformalGauge}. 

In the TPS, Legendre transformations are described as a set of strict contactomorphisms, that is, transformations which leave the Gibbs fundamental (contact) one-form invariant, thus expressing in geometrical language the invariant description of thermodynamics under this set of transformations. This idea served as a motivation for the formulation of Geometrothermodynamics (GTD), whose basic premise is that the geometric description of thermodynamics must also be invariant under Legendre transformations. Consequently, the metric in the TES should be invariant under this set of transformations. Legendre transformations are naturally defined on the TPS; therefore, in the GTD formalism the metric on the TPS must be invariant. Moreover, the TES is considered as a subspace of the TPS in such a way that the existence of an invariant metric for the TES is guaranteed which, in turn, inherits the Legendre invariant property of the TPS metric \cite{2007Quev,2008Queva,2011Quevb,2009Quevb}.

A geometric theory of thermodynamics in the TES in terms of a Riemannian structure naturally entails the idea of an invariant description under re\-pa\-ra\-me\-tri\-za\-tions of the intensive and/or extensive variables used to describe the system. In this work, we are interested in analyzing how such reparametrizations can be described in the TPS from a geometric point of view. To this end, we consider here reparametrizations as determined by mappings that are $C^\infty$ functions of their corresponding parameters. It is important to remark the physical motivation for considering this analysis. The kind of physical situation that these reparametrizations aim to describe arises from the possibility of not always being able to control the intensive variables of a thermodynamic system in contact to some reservoirs, and/or not having a direct access to the other thermodynamic variables by means of measurements, but only through some general functions of the original state variables instead. 
From the point of view of a geometric description of thermodynamics point, it is interesting to analyze the consequences that such physical situations could have on the geometric structures of both physical manifolds, namely the TES and the TPS.

As a result of these investigations we find that different contact and Riemannian structures of the TPS can be considered to account for the reparametrizations, while the Riemannian structure of the TES  remains unaltered.
That is, for each different parametrization, the TES, as a Riemannian manifold, can be embedded in a TPS with a different contact structure and metric. In particular, we find that the information of the reparametrizations on the TPS is codified in a rank-two tensor on this space. Nonetheless, under the corresponding pullbacks such different geometric structures on the TPS are mapped to the same Riemannian submanifold of equilibrium states. In this sense, we have different TPS's for the same TES, depending on the parametrization chosen for the thermodynamic variables.  

The structure of the paper is as follows. In Section \ref{sec:sta}, we review on how the Boltzmann-Gibbs exponential distribution can be obtained using the optimized Lagrange multipliers method (OLM) for maximizing the Boltzmann entropy and explore how some quantities transform under reparametrizations of the corresponding intensive variables. Moreover, we compare the concept of reparametrizations, as used in statistical approaches, with the more strict concept of diffeomorphism and coordinate transformations, as used in geometry. In Section \ref{sec:fluctuations}, we review the link between statistical fluctuations and the geometry of the TES. We show explicitly that the reparametrizations we define on the TPS correspond to coordinate transformations on the TES. In Section \ref{sec:contact}, we analyze how the reparametrizations can be introduced in the TPS through different contact and Riemannian structures and show that, despite having different TPS's, the Riemannian structure of the TES remains unchanged. Finally, in Section \ref{sec:conclusions} a discussion of the results found in this work is presented.

\section{Statistical reparametrization}
\label{sec:sta} 

In this section we begin with a review of the maximum entropy principle using the optimized Lagrange multipliers method to find the Boltzmann-Gibbs distribution. Then, we introduce a first notion of reparametrization for the TES and state some results on the Hessian matrix and the Fisher matrix (borrowing the terminology from information theory) that will be useful in the following sections. The OLM method was used in the work of Mrugala et al.\cite{1990Mrug}, where a link between statistical mechanics and a geometric description of thermodynamics was established in terms of a Riemannian contact manifold. The underlying idea is the interpretation proposed by Jaynes of viewing statistical mechanics as a theory of maximum entropy inference \cite{1957Jayn}. From a mathematical point of view, this interpretation renders statistical inference almost indistinguishable from (subjective) statistical mechanics. This point of view allows one to trade physical assumptions for statistical inference based on the least biased information input. 
This approach was retaken\cite{1990Mrug} to establish a link between statistics and the geometric structures of thermodynamics.

Thus, we briefly review the procedure given by Mrugala et al. \cite{1990Mrug}. Let us consider a (mechanical) phase space $\Gamma$ along with a set of $n$ functions $H_a: \Gamma \to \mathbb{R}$ where $a=1,\ldots,n$. Given the set of statistical averages 
$\{\langle H_a \rangle \}$, the task is to find the probability distribution $\rho: \Gamma \to \mathbb{R}^+$ which maximizes the Boltzmann entropy 
	\be \label{shannon}
	S[\rho] = -\int \rho\ \ln\!\ \rho \ dx\,,
	\ee
subject to the following set of $n+1$ constraints, 
	\begin{align}  \label{constraints}
	1 &= \int \rho\ dx\,, \nonumber \\
	E_a &= \langle H_a \rangle = \frac{\int H_a \rho \ 
	dx}{\int \rho \  dx}\, ,
	\end{align}
where each $E_a$ is associated to the statistical ensemble average of the corresponding microscopic functions $H_a(x)$ and $x \in \Gamma$. The OLM method allows us to maximize the entropy functional \eqref{shannon} subject to the set \eqref{constraints} of $n+1$ constraints. Then, let us consider the following functional 
	\be \label{entropyfunctional}
	\mathcal{I}[\rho] = -\int \rho \ln \rho \ d x - 
	\phi \left[\int \rho \ d x - 1\right] + I^a 
	\left[\int H_a \rho \ d x  - E_a \right]\,,
	\ee
where $\phi$ is the Lagrange multiplier associated to the first constraint, often known as the normalization condition which we will later identify as the total Massieu potential, and $\{I^a\}$ is a set of $n$ multipliers associated with the other $n$ constraints which we will eventually use to parametrize the TES manifold. 

As is well known, the result of extremizing the entropy functional \eqref{entropyfunctional} is the exponential distribution given by
	\be \label{genrho}
	\rho \left(\Gamma; \phi, I^1, \ldots, I^n \right) = \exp \left(-\phi + I^a H_a \right)\,,
	\ee
where a summation over $a$ in $I^a H_a$ is implicit. Throughout this work, we will adopt the convention that any set of repeated indices indicates a sum over all its values, unless stated otherwise.

It is obvious that applying the same procedure using a set of $m$ functions $H_\mu: \Gamma \to \mathbb{R}$, such that 
$\{H_\mu\} \subset \{H_a\}$ where $m < n$, we obtain a different exponential distribution characterized by $m$ parameters $\{I^\mu\}$ 
and $\bar\phi$, 
and where $E_\mu = \langle H_\mu \rangle$ with respect to
	\be \label{canondist}
	\bar{\rho} = \exp ( -\bar{\phi} + I^\mu H_\mu )\,.
	\ee
The resulting distributions $\rho$ and $\bar{\rho}$ are different and so are the corresponding statistical averages $E_a$ and 
${E}_\mu$. In the context of information geometry, there is a similar notion where statistical estimation is done on submanifolds of the statistical manifold representing a family of curved exponential distributions \cite{amaribook}.

The parameter $\phi$ can be obtained as a function of the parameters $\{I^a\}$ by imposing the first condition in Eq.\,\eqref{constraints},
	\be \label{wtilde}
	\phi(I) = \ln \int \exp(I^a H_a)\ d x = \ln Z(I)\,,
	\ee
where 
	\be \label{Z}
	Z \equiv \int \exp(I^a H_a) \  d x \,
	\ee
is the partition function of the system, which is a function of the parameters $\{I^a\}$. We identify the parameters $\{I^a\}$ with a set of intensive thermodynamic variables and $\phi(I)$ is nothing but the total Massieu \textcolor{red}{(molar or volumetric)} potential. \textcolor{red}{We must remark that for a homogeneous (degree one) closed system, one must consider the molar or volumetric thermodynamic variables; otherwise the thermodynamic potential is identically zero. See, for instance, \cite{1995Rupp} for the description of a set of thermodynamic metrics in terms of volumetric variables. This is not the case for other degrees of homogeneity or even for non-homogeneous systems; therefore, we will refer to the variables generically as intensive and extensive.} Hence the extensive variables are given as
	\be \label{eqstate}
	\langle H_a \rangle  = E_a = \frac{\partial \phi}{\partial I^a}.
	\ee
	
Then, we can introduce a first notion of reparametrization as follows. Consider another set of parameters $\{I^{\tilde a}\}$ related by an invertible mapping with the intensive variables $\{I^a\}$, that is, $\{I^{\tilde a} \} = \{I^{\tilde a} (I)\}$ and the transformation
matrix 
\be
 \Lambda^{\tilde a}{}_a = \frac{\partial I^{\tilde a}}{\partial I^a}\ 
\label{reparametrize}
\ee
has non-vanishing determinant. 

Clearly, the statistical average can be expressed  as
	\be \label{fa}
	\langle H_{\tilde a} \rangle = E_{\tilde a} =  \Lambda^a{}_{\tilde a} \frac{\partial \phi}{\partial I^a} \, ,
	\ee
where the subindex in $\langle H_{\tilde a} \rangle$ indicates that the average is a function of the parameters $\{I^{\tilde a}\}$. 

We note that while $\phi$ and $\rho$ behave as scalar functions under the introduced reparametrization \eqref{reparametrize}, the statistical averages transform as the components of a one-form, see \eqref{fa}; thus we will drop the tilde notation from $\rho$ and $\phi$ altogether. In section \ref{sec:contact} we will analyze these transformation properties from a geometrical point of view and we will introduce another notion of reparametrization related to $\{E_a\}$ pertaining to a higher dimensional space.

We close this section commenting on the idea of the reparametrizations considered here from a geometric point of view, with reference to the contents of the following sections. To this end, consider the TES of a physical system as an $n$-dimensional differential manifold $\mathcal{E}$ parametrized by a set of thermodynamic variables which can be either extensive or intensive, depending on the chosen thermodynamic potential. For definiteness, we have chosen the set of $n$ parameters serving as coordinates for $\mathcal{E}$ as the intensive variables $I^a$. The reparametrizations here introduced might be considered, from a geometrical point of view, as a change of coordinates in the thermodynamic manifold $\mathcal{E}$.
\textcolor{red}{Nevertheless, we must mention that the reparametrizations introduced in this work will be considered as diffeomorphisms on the TPS. From the geometric point of view, on the TPS all the variables are independent and it is only on the TES that they acquire a thermodynamic character. }

Furtheremore, in section \ref{sec:contact} we give an explicit and formal definition for the TES, the TPS and reparametrizations in terms of contact contact manifols and its coordinates.

\section{The geometry of fluctuations}
\label{sec:fluctuations}

In this section, we review the link that exists between the fluctuations of a quantity usually referred in literature as, the microscopic entropy and thermodynamic geometry \cite{1990Mrug}. Simultaneously, we also verify that the reparametrizations introduced in the preceding section actually behave as a coordinate transformation on the $\mathcal{E}$ manifold.

In the previous section, a maximum entropy principle was considered to find a probabilistic distribution; in that case, the relevant quantity from a physical and statistical mechanical point of view was the Boltzmann entropy given by Eq.\,\eqref{shannon}. The functional $S$ can be understood as the average with respect to the corresponding distribution of the microscopic entropy, 
\be \label{micros}
	s = - \ln \rho\,,
\ee
which implies 
\be \label{averageS}
	S = \langle s \rangle = - \int \rho \ln \rho\ d x\,.
\ee
If the normalization constraint is not imposed on the distribution, the microscopic entropy can be considered as a function of the coordinates of $\Gamma$ and of the set of $n+1$ parameters, $s = s(x; \phi, I^a)$,  whose explicit expression is
\be  \label{s}
	s = \phi - I^a H_a(x)\,.
\ee
We can immediately see that the microscopic entropy $s(x;I^a)$ behaves as a scalar function with respect to the intensive parameters $\{I^a\}$. Therefore, considering the reparametrizations introduced in the previous section we observe that  
\be \label{stilde}
	s = \phi - I ^{\tilde a} H_{\tilde a}(x) \,
\ee
describes the same microscopic physical information.

If we consider $s$ as a function on the ($n+1$)-dimensional space, its differential is given by 
	\be \label{ds}
	ds = d\phi - H_a(x) \ dI^a \, ,
	\ee
which can be expressed in terms of the $I^{\tilde a} $ coordinates as 
\be \label{dstilde}
	ds = d\phi - \Lambda^{\tilde a}{}_a H_{\tilde a}d I^a \,.
\ee

We are interested in the average and the variance of the differential of the microscopic entropy $\d s$ with respect to $\rho$, as they are related to the contact and Riemannian structures of the TPS introduced in the next section. The average value of the differential of the microscopic entropy considering the reparametrization is
\be \label{meandstilde}
\langle ds \rangle = d\phi - \langle H_{\tilde a} \rangle{} \Lambda^{\tilde a}{}_a  dI^a \,,
\ee
while its variance $\langle (d s - \langle d s \rangle)^2 \rangle$ can be used to define a metric $g$ on the TES  

\begin{equation}\label{vardstilde}
	g = \langle (d s - \langle d s \rangle)^2 \rangle = \Lambda ^{\tilde a}{}_a \Lambda^{\tilde b}{}_b \ 
	\langle (H_{\tilde a} - \langle H_{\tilde a} \rangle)(H_{\tilde b} - \langle H_{\tilde b} \rangle) \rangle \ d I^a \otimes d I^b \ ,
\end{equation}
which can be expressed in terms of the ${I^{\tilde a}}$ coordinates as, 
\begin{equation}\label{vardstilde2}
g=	\langle (H_{\tilde a} - \langle H_{\tilde a} \rangle)(H_{\tilde b} - \langle H_{\tilde b} \rangle) \rangle \ d I^{\tilde a} \otimes d I^{\tilde b} \, .
\end{equation}

In terms of the original intensive variables, the results in Ref.\cite{1990Mrug} are recovered, 
\be \label{canonmeands}
	\langle ds \rangle = d\phi - \langle H_a \rangle\ dI^a\,,
\ee
with the variance and the metric as 
	\be  \label{canonvards}
	g= \langle (d s - \langle d s \rangle)^2 \rangle = \langle 
	(H_a - \langle H_a \rangle)(H_b - \langle H_b \rangle) 
	\rangle \ d I^a \otimes d I^b \,.
	\ee
We recognize the components in Eq.\,\eqref{canonvards} as the variance-covariance matrix for $\{H_a\}$ \cite{crooks}, and from \eqref{vardstilde} we note that under the reparametrization the metric transforms accordingly. Therefore, this metric contains the information about statistical fluctuations for the Boltzmann-Gibbs distribution. This metric can be utilized to define a Riemannian manifold 
$(\mathcal{E},g)$ where $\mathcal{E}$ is the TES. 

The components of the metric \eqref{canonvards} can also be recognized as the Fisher information matrix in the context of information geometry defined as

	\begin{align} \label{genFisher}
	g_{ab} &= \bigg \langle \frac{\partial \ln \rho}{\partial I^a} 
		\ \frac{\partial \ln \rho}{\partial I^b} \bigg\rangle \nonumber \\
	&= \Lambda^{\tilde a}{}_a \Lambda^{\tilde b}{}_b  \ \bigg \langle \frac{\partial \ln \rho}{\partial 
		I^{\tilde a}} \ \frac{\partial \ln \rho}{\partial I^{\tilde b}} \bigg
		\rangle  \nonumber \\
	&= \Lambda^{\tilde a}{}_a \Lambda^{\tilde b}{}_b  \ g_{\tilde a \tilde b} \, ,
\end{align}
where $\rho$ is the exponential distribution given by \eqref{genrho}. From \eqref{genFisher}, it can be verified that the Fisher matrix components indeed transform as the components of a metric on $\mathcal{E}$. It was noticed in \cite{1945Rao} that \eqref{genFisher} can be used to define a metric on a statistical manifold. In the context of thermodynamics, (minus) the Hessian of the entropy has been used as the components for a metric defined on the space of equilibrium states $\mathcal{E}$ \cite{1979Rupp, 1990Mrug}. The reparametrizations  introduced here will be considered, from a geometrical point of view, as a change of coordinates in the thermodynamic manifold $\mathcal{E}$.

Additionally, it can be shown that in this coordinate system $\{I^a\}$, the components of the metric $g$ can be expressed, for the distribution function \eqref{genrho}, in terms of the Hessian of the potential $\phi$, 

\be \label{hessianphi}
	g_{ab} = \frac{\partial^2 \phi}{\partial I^a \partial I^b}\,.
\ee 

However, this last form for the components of the metric is not, in general, a covariant expression, since it is written as the second partial derivative of a scalar function. 
Nevertheless, equation \eqref{hessianphi} is the most common expression found in the literature of thermodynamic geometry  for the components of the thermodynamic metric. 

\section{Reparametrizations of the thermodynamic phase space}
\label{sec:contact}

In the previous sections, it was shown that statistical reparametrizations of the form $I^a \rightarrow I^{\tilde a} (I)$ can be considered as coordinate transformations in the  equilibrium space 
${\mathcal E}$. In the description of this space, 
the dependent variables $\phi(I)$ and $E_a(I)$ also play an important role. 
In this section, we consider the entire set $\{\phi, E_a, I^a\}$ as independent variables and use them as coordinates for the TPS. We also introduce a second notion of reparametrization pertaining to the TPS and 
analyze its contact and Riemannian structures by extending the previously introduced reparametrization in $\mathcal{E}$ to this space. We also single out a tensor field which contains the information about the reparametrizations of the variables in the TPS. 

We  define the TPS as a $(2n + 1)$-dimensional Riemannian contact manifold $(\mathcal T, 
{\eta}, {G})$ with a contact one-form ${\eta}$ and a Riemannian metric $G$ 
from which the metric \eqref{vardstilde} can be obtained as the pullback induced by the embedding map ${\varphi}: \mathcal{E} \to \mathcal T$, where $\mathcal{E}$ is the TES. In this case, the TES is given as a Legendre submanifold $\mathcal{E} \subset \mathcal T$ which is an integral submanifold of $\mathcal T$ of maximal dimension, i.e., $\mbox{dim}(\mathcal E)=n$. Therefore, we give the following local characterization of $\mathcal{E} \subset \mathcal T$ as

\be 
\mathcal E:=\left\lbrace \left\lbrace \phi , E_a , I^a \right\rbrace \in \mathcal T \mid \d\phi -E_a\d I^a=0 \quad \text{and} \quad \Phi(I^a)=\ln{\int \exp{\left( I^a H_ a\right) }\d x}    \right\rbrace . \nonumber
\ee
 The conceptual details of the construction of Riemannian contact manifolds and its Legendre submanifold are given in the appendix. 

At this point, we would like to introduce a second notion of reparametrization. Recall that under the change of coordinates 
$I^{\tilde a} = I^{\tilde a}(I)$ of ${\mathcal{E}}$, the statistical ensemble averages $\langle H_a \rangle = E_a$ behave as the components of the one-form $d\phi$. In the $(2n +1)$-dimensional TPS manifold, $\{\phi,E_a,I^a\}$ are considered as coordinates, so there is no relation between the parameters $\{I^a\}$ and $\{E_a\}$. Once we determine a particular embedding, we find the relation \eqref{eqstate} between the thermodynamic state variables, i.e., the parameters in the TPS acquire a physical meaning as the intensive thermodynamic variables $\{I^a\}$, the extensive ones $\{E_a\}$ and the thermodynamic potential $\phi(I)$, recognizing \eqref{eqstate} as the equations of state. Therefore, it is in the TPS where we can introduce a notion of reparametrization as a diffeomorphism $\Psi: \mathcal{T} \to \mathcal{T}$. As a consequence of the application of the diffeomorphism $\Psi$, we use two different set of coordinates to describe the TPS, namely, $\{\phi, E_a, I^a\}$ and $\{\tilde{\phi}, E_{\tilde a}, I^{\tilde a}\}$. According to Darboux theorem (see the appendix), the one-forms
\be
\eta_1 = d\phi - E_a d I^a \ , \quad \eta_2 = d\tilde{\phi} - E_{\tilde a} d I^{\tilde a}
\ee
can be used to introduce a contact structure in $\mathcal{T}$. We will restrict ourselves to a particular set of reparametrizations. In the TPS we consider diffeomorphisms such that $\{E_{\tilde a} \} = \{E_{\tilde a}(E)\}$, that is, each new variable $E_{\tilde a}$ depends only on the $\{E_a\}$ variables. Furthermore, extending the reparametrizations of the previous section, $\{I^{\tilde a} \} = \{I^{\tilde a}(I)\}$, to the TPS, we will understand a reparametrization of the state variables as a diffeomorphism $\Psi: \mathcal{T} \to \mathcal{T}$ given as $\{\tilde{\phi}, E_{\tilde a}(E), I^{\tilde a}(I)\}$, restricted to the particular cases with $\tilde{\phi}(\phi,I,E)=\phi$. By construction, $\Psi$ it is not a contact transformation. Then, it follows that $\eta_1$ and $\eta_2$ are not equivalent.

Moreover, we can introduce two different bilinear symmetric tensor fields
\be \label{tensort}
	t_1 = d E_a {\otimes} d I^a \ , \quad
	t_2 = d E_{\tilde a } {\otimes} d I^{\tilde a }\ .
\ee

As such, the symmetric tensors \eqref{tensort} cannot be considered as metrics for $\mathcal T$ since they are degenerate. However, it is possible, as done in Ref.\cite{1990Mrug}, to remedy this flaw by adding the tensor product of ${\eta _1}$ and $\eta _2$ to $t_1$ and $t _2$ respectively to obtain 
\begin{equation} \label{2metricG}
	{G_1} = {\eta_1} \otimes {\eta_1} + {t_1} \, , \quad
	{G_2} = {\eta_2} \otimes {\eta_2} + {t_2} \ .\end{equation}
These metrics can also be written in terms of the local coordinates $\{\phi,E_a,I^a\}$  as 
\be \label{GFRcoord}
	G_1 = (d \phi - E_a \ d I^a) \otimes (d \phi - E_b d I^b) + 
	d E_a {\otimes} d I^a\,.
\ee
\be \label{G2}
	{G_2} = \bigg(d\tilde{ \phi} - E_{\tilde a} \frac{\partial I^{\tilde a}}{\partial I^a} d I^a \bigg) \otimes \bigg(d \tilde{ \phi} - E_{\tilde b} \frac{\partial I^{\tilde b}}{\partial I^b} d I^b \bigg) \ + \ \frac{\partial I^{\tilde a}}{\partial I^a} \frac{\partial E_{\tilde a}}{\partial E_b} \ d E_b {\otimes} d I^a \, .
\end{equation}

The Legendre submanifold $\mathcal{E}$ is determined 
by the embedding $\varphi: \mathcal{E} \to \mathcal T$ and the condition $\varphi^*(\eta_1) = 0$, which describes geometrically the first law of thermodynamics and the equations of state expressed in terms of the total Massieu potential $\phi$. 
Using the coordinates $\{\phi,E_a, I^a\}$, this embedding yields the equations 
\be \label{firstlaw} 
	d \phi = E_a d I^a \quad \text{and} \quad E_a = 
	\frac{\partial \phi}{\partial I^a}\,.
\ee 
On the other hand, in the local coordinates $\{\phi,E_{\tilde a}, I^{\tilde a}\}$, the embedding
which preserves the first law of thermodynamics is determined by the condition 
$\tilde\varphi^*(\eta_2) = 0$, i.e.
\be 
\label{embedding}
	d \tilde{ \phi}= E_{\tilde a} d I^{\tilde a} = E_{\tilde a} 
	\frac{\partial I^{\tilde a}}{ \partial I^a} dI^a\ , \quad
	E_{\tilde a} = \frac{\partial\phi}{\partial I^a}\frac{\partial I^a}{\partial I^{\tilde a}}\ .
\ee

The pullback $g=\varphi^*(G_1)$ induced by $\varphi$ yields the metric \eqref{canonvards} (or in terms of the coordinate system $\{I^a\}$ as Hessian metric  components \eqref{hessianphi}) on 
$\mathcal{E}$, cf. Eq.\,(16) in Ref. \cite{1990Mrug} and Eq.\,(25) in Ref. \cite{HamThermo} (note that our convention differs from that of Ref.\cite{HamThermo} by a sign). 
From the pullback of the metric \eqref{G2} induced by the mapping $\tilde{\varphi}: \mathcal{E} \to \mathcal T$ 
we obtain Eq.\,\eqref{vardstilde} (or Eq.\,\eqref{genFisher} in terms of the coordinate system $\{I^a\}$), that is, the same metric $g_{ab}$ on $\mathcal{E}$, just written in the reparametrized variables. Indeed, using \eqref{embedding} we obtain
\begin{align} \label{1metricg}
	\tilde{\varphi}^* ({G_2}) &= \frac{\partial^2 \tilde{ \phi}}{\partial I^{\tilde a} \partial I^{\tilde b}}\ d I^{\tilde a} \otimes d I^{\tilde b} \nonumber \\ 
	&= \langle (H_{\tilde a} - \langle H_{\tilde a} \rangle)(H_{\tilde b} - \langle H_{\tilde b} \rangle)\rangle\ d I^{\tilde a} \otimes d I^{\tilde b} \nonumber \\
	&= \Lambda^a{}_{\tilde a} \Lambda^b{}_{\tilde b}\ \langle (H_a - \langle H_a \rangle)(H_b - \langle H_b \rangle)\rangle\ d I^{\tilde a} \otimes d I^{\tilde b} \,  \\
	&= \langle (H_a - \langle H_a \rangle)(H_b - \langle H_b \rangle)\rangle\ d I^a 
	\otimes d I^b \nonumber \, \\
	&= \varphi ^*(G_1)\nonumber,
\end{align}
where \eqref{vardstilde} and \eqref{genFisher} were considered in order to arrive at the last result. Thus we have obtained the same induced metric \eqref{canonvards} for $\mathcal{E}$ making the pullback of a different Riemannian structure on $\mathcal{T}$.

Finally, let us consider a particular set of reparametrizations, 
$I^{\tilde a} (I) = I^{\tilde a}(I^a)$ and $E_{\tilde a} (E) = E_{\tilde a}(E_a)$, that is, $\frac{\partial I^{\tilde a}}{\partial I^a} = 0$, unless $\tilde{a} = a$ and similarly for $E_{\tilde a}$. For instance, if for the intensive variables we consider the temperature and pressure, $I^a = \{T, p\}$ a reparametrization of this kind would be $\{\tilde{T} = f^1(T), \tilde{p} = f^2(p)\}$ where $f^1$ and $f^2$ are smooth functions. In this case the metric in the TPS given by \eqref{G2} takes the corresponding form to this particular set of restricted diffeomorphisms
	\begin{equation}	
	\tilde{G} = \bigg(d \tilde{ \phi} - \sum_{a, \tilde{a}} E_{\tilde a} \frac{d I^{\tilde a}}{d I^a} d I^a \bigg) \otimes \bigg(d \tilde{ \phi} -  \sum_{b, \tilde{b}} E_{\tilde b} \frac{d I^{\tilde b}}{d I^b} d I^b \bigg) \ + \ \sum_{a, \tilde{a}} \frac{d I^{\tilde a}}{d I^a} \frac{d E_{\tilde a}}{d E_a} \ d E_a {\otimes} d I^a \, ,
	\end{equation}
where we have written explicitly the sums over the indices to avoid any confusion arising from the notation for this particular case.

\textcolor{red}{As a final remark on the diffeomorphisms introduced here, we must say that no further restrictions are imposed on them besides their specific form, namely, $\Psi: \mathcal{T} \to \mathcal{T}$ with $\{\tilde{\phi}, E_{\tilde a}(E), I^{\tilde a}(I)\}$ and the condition $\tilde{\phi}(\phi,I,E)=\phi$. It could be thought that for homogeneous systems of degree one these mappings could alter the extensive nature of the variables, but as we have shown, the geometric thermodynamic description on the TES remains the same. Therefore, these reparametrizations do not affect the extensive or non-extensive nature of the system.} 


\section{Conclusions}
\label{sec:conclusions}

Let us summarize the results that we have found in this work and single out some of their consequences. We start with the Riemannian manifold $(\mathcal{E},g)$ with the metric $g$ given by \eqref{hessianphi}. This is an embedded manifold in a higher-dimensional manifold possessing a contact structure and a metric $(\mathcal T,\eta _1, G_1)$. In physical terms, these correspond to the TES and the TPS, respectively. In Ref.\cite{1990Mrug} it is shown that under a total Legendre transformation the metric \eqref{GFRcoord} gives rise, after the pullback, to Ruppeiner's metric \cite{1979Rupp} on the space of equilibrium states $\mathcal{E}$.  

We have seen that under a reparametrization $I^{\tilde a} = I^{\tilde a}(I)$ we can describe the manifold $(\mathcal{E},g)$ in terms of
another TPS, $(\tilde{\mathcal T}, \eta_2, G_2)$. We identify this as a different TPS because the contact one-form $\eta_1$ is not related to $\eta_2$ via a contact transformation. Nevertheless, the thermodynamic description on $\mathcal{E}$ is the same, since the first law \eqref{firstlaw} is recovered from the condition ${\varphi}^*({\eta_1}) = 0$, as can be verified from the corresponding embedding conditions. Moreover, the metric on the TES, which is related to the statistical 
(thermodynamic) fluctuations, is obtained as $g = {\varphi}^*({G_1})$ or 
$g = {\tilde\varphi}^*({G_2})$. 

Finally, it was found that the pullbacks of the two different metrics in the TPS can be expressed only in terms of the tensors $t_1$ and $t_2$, that is, $g = \tilde{\varphi}^*({t_1})$ and 
$g = \varphi^*(t_2)$, respectively. In Ref.\cite{1990Mrug} it is argued that the contact structure describes the thermodynamic equilibrium, whilst the metric structure describes the thermodynamic fluctuations. 
As we have seen, both contact one-forms, $\eta _1$ and $\eta _2$, lead to the first law of thermodynamics on $\mathcal{E}$ and the two tensors $t_1$ and $t_2$ give the same geometric description of the thermodynamic fluctuations on $\mathcal{E}$ through the metric $g$. Therefore, despite having different TPS's via these particular reparametrizations of the intensive and extensive thermodynamic variables, the geometric structures in the thermodynamic state space remain invariant. This indicates that it is  possible to geometrically characterize the reparametrizations introduced here in the TPS by modifying the corresponding contact and Riemannian structures.

Thus, we can conclude that the tensor field $t_2$ geometrically comprises the information of these particular reparametrizations of the thermodynamic state variables in the description of statistical fluctuations in the TPS. Then, we see that geometrically speaking the proposed reparametrizations have no effect on the TES, that is, the corresponding system remains in equilibrium. 
However, it could be interesting to analyze the consequences of such reparametrizations for the geometric description of thermodynamic processes near  equilibrium, which must be described by curves outside the TES \cite{HamThermo}. It could also be attempted to construct a TPS with the contact structure  $\eta _1$ and a metric resulting from combining the canonical one-form with the tensor field $t_2$. 
This would allows us to analyze the geometric information about reparametrizations inherited on the TES  to single out the effects that these reparametrizations have on the description of fluctuations. These investigations have been developed in Ref.\,\cite{Pineda2019}, where a conection between these ideas and a relation of Legendre invariant metrics and fluctuations of thermodynamic variables were explored.

\section*{Acknowledgments}

L. F.  Escamilla-Herrera and F. Nettel acknowledge  financial support from the Consejo Nacional de Ciencia y Tecnolog\'\i a 
(CONACyT, M\'exico). C. Gruber acknowledges support by a Junior Fellowship of the Hanse-Wi\-ssen\-schafts	-ko\-lleg Delmenhorst, and from the University of Oldenburg and the Research Training Group "Models of Gravity". This work was partially supported  by UNAM-DGAPA-PAPIIT, Grant No. 111617 and by the Ministry of Education and Science of RK, Grant No. BR05236322 and AP05133630. V. Pineda-Reyes acknowledges financial support from CONACyT, M\'exico through project A1-S-31269 by Grant No. 29849.
The authors thank Cesar Lopez-Monsalvo for fruitful discussions.

\appendix

\section{Contact manifolds}


In this appendix, we briefly review the main properties of contact manifolds. Consider a $(2n +1)$-dimensional differential 
manifold $\mathcal T$. A contact structure is a maximally non-integrable distribution $D \subset \mathrm{T} \mathcal T$ of $2n$-dimensional hyperplanes. Such a structure is characterized by means of the equivalence class of 1-forms $[\eta]$ such that for any representative $\eta$ the relation 
\be  \label{dist}
	D = \mathrm{ker}(\eta) 
\ee
is satisfied, and the condition of non-integrability 
\be \label{inte}
	\eta \wedge (\eta)^n \neq 0
\ee
is fulfilled. The equivalence class $[\eta]$ is defined by a conformal relation, that is, $\eta$ and $\eta'$ are equivalent if $\eta = \Omega \eta'$, where $\Omega$ is a non-vanishing real function. The condition \eqref{inte} means that a well-defined volume form exists on $\mathcal T$. Introducing a set of local coordinates $\{\phi, x_a, y^a\}$ it is possible to express $\eta$ in its canonical (Darboux) form,
\be \label{etadarboux}
	\eta = d \phi - x_a d y^a\,.
\ee
A transformation $f: \mathcal T \to \mathcal T$ preserving the contact structure, $f^* (\eta) = \Omega \eta$, is called a contact transformation. A  (discrete) Legendre transformation leaves invariant the contact form, $f^* (\eta) = \eta$, i.e., it represents a symmetry of the contact form and is defined as follows. Let $I\cup J$ be any partition of a set of $n$ indices into two disjoint sets $I$ and $J$, then a Legendre transformation on $\mathcal T$ is given by the following $2n + 1$ equations between the sets of coordinates $\{\phi,x_a,p^a\}$ and $\{\phi',x'_a,y'^a\}$, 
\be \label{legendretransf}
	\phi' = \phi -  y^i x_i, \quad y'^i = -x_i, \quad y'^j = y^j, 
	\quad x'_i = y^i, \quad x'_j = x_j\,,
\ee
with $i \in I$ and $j \in J$. A direct calculation shows that 
\be \label{etacov}
	\eta = d \phi - x_a \ d y^a = d \phi' - x'_a d y'^a\,. 
\ee
For $J = \emptyset$, we say that $f$ is a total Legendre transformation, otherwise we have a partial Legendre transformation.

A Legendre submanifold is a maximally integral embedded submanifold $\mathcal{E} \subset \mathcal T$ whose tangent bundle is completely contained in the distribution $D$, $\mathrm{T}\mathcal{E} \subset D$. A characterization in terms of local coordinates is given by the following theorem \cite{2013Arno}. Let $I\cup J$ be the same partition as before and consider any function $\Phi = \Phi(x_i,y^j)$ where $i \in I$ and $j \in J$. The following set of $n+1$ equations defines a Legendre submanifold $\mathcal{E}$ of $\mathcal T$,
\be \label{legendresub}
	 x_j = \frac{\partial \Phi}{\partial y^j}, \quad y^i = 
	 -\frac{\partial \Phi}{\partial x^i}, \quad \phi = \Phi + 
	 x_i \frac{\partial \Phi}{\partial x_i} \,.
\ee
Conversely, any Legendre submanifold is defined locally by these equations for at least one of the $2^n$ partitions of the set $I \cup J$. Such a maximally integrable submanifold is of dimension $n$ and is determined by the condition $\varphi^*(\eta) = 0$, where $\varphi: \mathcal{E} \to \mathcal T$ is the embedding mapping. 

Finally, we can furnish the contact manifold with a Riemannian structure, defining a metric $G$ on $\mathcal T$. Then,  
$\mathcal T$ becomes a Riemannian contact manifold $(\mathcal T, \eta, G)$. Moreover, a canonical metric $g$ is induced on
the submanifold $\mathcal E$  by means of $g = \varphi^*(G)$.  

\bibliographystyle{unsrtnat}

\bibliography{G3}

\end{document}